\documentclass[fleqn,twoside]{article}

\usepackage{amsmath}
\usepackage{espcrc2}
\usepackage{graphicx}

\newcommand{\keV}{\ensuremath{\:\text{keV}}}

\newcommand{\GeV}{\ensuremath{\:\text{GeV}}}
\newcommand{\TeV}{\ensuremath{\:\text{TeV}}}
\newcommand{\cm}{\ensuremath{\:\text{cm}}}
\newcommand{\mtr}{\ensuremath{\:\text{m}}}
\newcommand{\km}{\ensuremath{\:\text{km}}}
\newcommand{\yr}{\ensuremath{\:\text{yr}}}

\title{SUSY at the Pole}

\author{J.\ Kersten\address[ICTP]{The Abdus Salam ICTP,
        High Energy Physics Section,
        Strada Costiera 11, 34014 Trieste, Italy}%
        \thanks{E-Mail:\texttt{jkersten@ictp.it}}}

\begin{document}

\begin{abstract}
We study the role neutrino telescopes could play in discovering
supersymmetric extensions of the Standard Model with a long-lived stau
next-to-lightest superparticle.  In such a setup, pairs of staus are
produced by cosmic neutrino interactions in the Earth matter.  In
optimistic scenarios, one can expect several pair events per year in a
cubic kilometre detector such as IceCube.  We also show that no
significant event rate can be expected for decays of staus stopped in
the detector.
\vspace{1pc}
\end{abstract}

\maketitle

\section{INTRODUCTION}

Cosmic neutrinos reach energies of at least $10^{11}\GeV$.  In
collisions with nucleons inside the Earth, the centre of mass energy
exceeds $1\TeV$ already for neutrino energies of about $10^6\GeV$.
Consequently, supersymmetric particles can be produced, provided that
SUSY exists close to the electroweak scale.  Hence, it is natural to ask
whether SUSY could be found in cosmic ray observatories.
Unfortunately, this is not possible in most scenarios, since the
produced superparticles immediately decay into the lightest one (LSP),
which is electrically neutral and thus not observable.

This problem can be avoided, if the next-to-lightest superparticle (NLSP)
is charged and long-lived.  This is possible, for example, if the LSP is
the gravitino, the superpartner of the graviton, and if $R$ parity is
conserved.  The most natural charged NLSP candidate in this case is the
lighter stau, usually composed predominantly of the scalar partner of
the right-handed tau.  Its decay length $L$ is roughly given by
\begin{equation}
\frac{L}{2 R_\oplus} \approx 
\bigg(\frac{m_{\tilde \tau}}{100\GeV}\bigg)^{-6}
\bigg(\frac{m_{3/2}}{400\keV}\bigg)^2
\bigg(\frac{E_{\tilde \tau}}{500\GeV}\bigg) ,
\end{equation}
where $m_{3/2}$ is the gravitino mass and $R_\oplus$ is the Earth
radius.  As we will only consider staus with energies above $500\GeV$,
they cross the whole Earth before decaying unless the gravitino is very
light.  Thus, we can treat the stau as a stable particle in the
following.

Cosmic neutrino interactions produce pairs of SUSY particles because of
$R$ parity conservation, which quickly decay into a pair of staus.  Due
to the large boost factor, these could then show up as upward-going,
nearly parallel
tracks in a neutrino telescope like IceCube \cite{Ahrens:2002dv}, as
suggested in \cite{Albuquerque:2003mi}.

\section{STAU PRODUCTION AND PROPAGATION}

The interactions leading to the production of superparticles are
analogous to the charged and neutral current neutrino-quark interactions
in the Standard Model (SM).  Instead of a $W$ or a $Z$, a chargino or
a neutralino is exchanged, resulting in a squark and a slepton.  We
calculated the cross section for two different SUSY mass spectra
\cite{Ahlers:2006pf}.  The first one is given by the benchmark point
corresponding to SPS $7$ \cite{Allanach:2002nj}.  The second one
(denoted by ``min $\widetilde m$'' in the following) consists of
squarks at $300\GeV$ and sleptons, charginos and neutralinos at
$100\GeV$.  In any case, the SUSY cross section is several orders of
magnitude smaller than its SM counterpart, mainly due to the much
heavier particles in the final state.

Travelling through the Earth, the staus lose energy chiefly due to
radiative processes at high energies.  The resulting energy loss scales
with the inverse particle mass, so that it is much smaller for staus
than for muons.  Hence, while muons have to be produced not more than a
few tens of kilometres outside the detector to be observable, staus are
visible even if produced much farther away
\cite{Albuquerque:2003mi,Reno:2005si}.  This may compensate for the
smaller production cross section.

\section{STAU DETECTION}

\begin{figure}
\includegraphics*[width=\linewidth,bb=0 0 385 370]{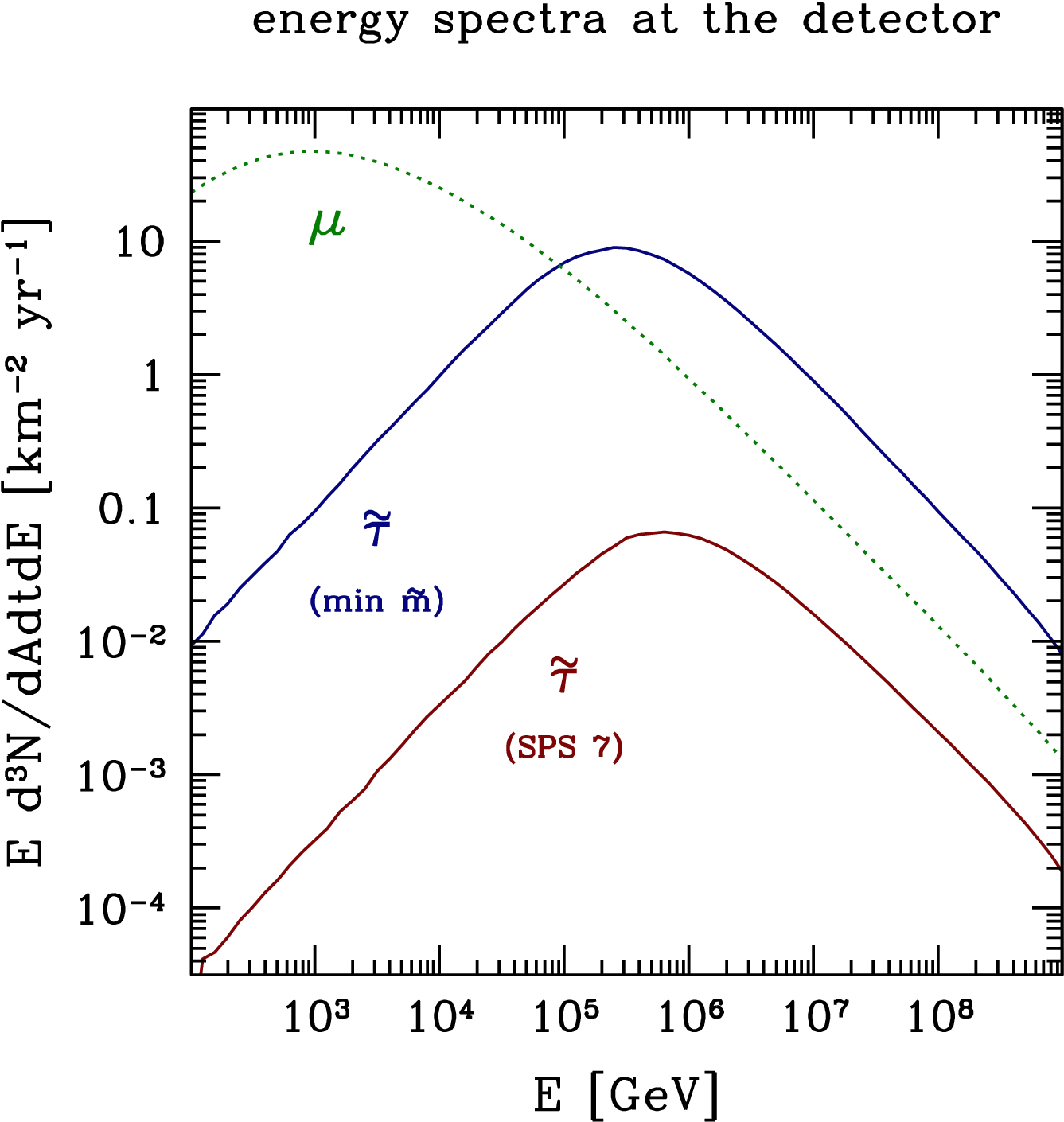}
\vspace{-1cm}
\caption{Fluxes of upward-going muons and staus
 for two different SUSY mass spectra (see text for details).
}
\label{fig:SingleSpectra}
\end{figure}

For the high-energy cosmic neutrinos, we assumed the Waxman-Bahcall flux
$E_\nu^2 \, F(E_\nu) \approx 
2\cdot 10^{-8}\cm^{-2}\:\text{s}^{-1}\:\text{sr}^{-1}\GeV$ per flavour
\cite{Waxman:1998yy}.  Note that current data allow for a flux larger by
about an order of magnitude \cite{Ribordy:2005fi}, so that we may have
underestimated the number of staus correspondingly.
Fig.~\ref{fig:SingleSpectra} shows the results for the spectra of muons
and staus, where an improved treatment of the energy loss at low
energies \cite{Ahlers:2006zm} was employed compared to
\cite{Ahlers:2006pf}.  Note that muons always dominate over staus
if one considers the spectrum in terms of the energy \emph{deposition}
in the detector, which is the actual observable.  Therefore, the total
rate of one-particle events can be used for reconstructing the neutrino
flux without taking into account the contribution from NLSPs
\cite{Ahlers:2006pf}.

\begin{figure}
\includegraphics*[width=\linewidth,bb=0 0 385 370]{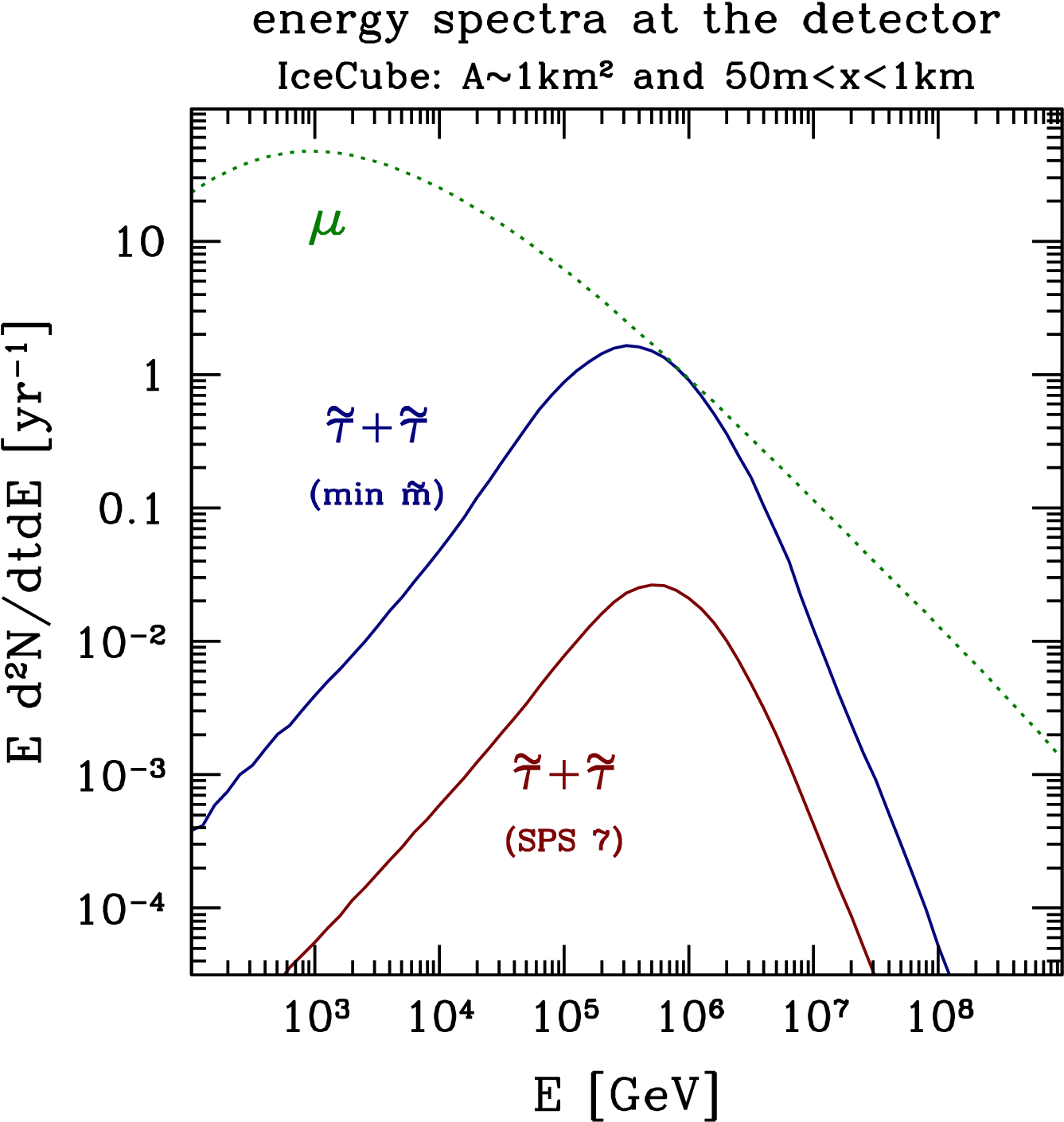}
\vspace{-1cm}
\caption{Rates of parallel stau tracks through a detector like IceCube
 for two different SUSY mass spectra (see text for details).
}
\label{fig:Spectra}
\end{figure}

IceCube will be able to identify the two tracks from a stau pair, if
their separation is greater than about $50\mtr$ and less than $1\km$
\cite{Ribordy:2006qd,SPIERING}.  Assuming that this separation can be approximately
calculated from the angle between the initial SUSY particles,
we obtained the rates of stau pairs shown in Fig.~\ref{fig:Spectra}.
The total number of stau pair events per year in
IceCube is about $5$ for the min $\widetilde m$ scenario.  For the SPS
$7$ mass spectrum, it decreases to $0.07$, chiefly due to the larger
squark masses and the accordingly smaller cross section for the
production of superparticles \cite{Ahlers:2006zm,Ahlers:2006pf}.

These numbers have to be compared to the background of muon pairs.
While the number of upward-going muons arriving at the same time just by
coincidence is tiny, a non-negligible number of muon pairs arises from SM
processes, for example if the initial neutrino-nucleon interaction
produces a muon and a hadron which decays to another muon.  However, as
muons have to be produced close to the detector, they will always be
separated by less than $50\mtr$ in IceCube and thus not contribute to
pair events \cite{Albuquerque:2006am}.

\section{STOPPED STAUS}

Staus with a very small energy will stop in the detector and decay after
a while, predominantly into a gravitino and a tau lepton.  The latter
can produce an observable cascade.  If this cascade can be correlated
with an upward-going particle track ending at the same position, which
may be possible in IceCube for stau lifetimes less than a few hours
\cite{SPIERING}, this provides another virtually background-free
signal that does not rely on the observation of stau pairs.  In addition,
the stau lifetime could be measured.  Let us therefore estimate the
rate of such events in the optimistic min $\widetilde m$ scenario, 
{\setlength{\arraycolsep}{2.5pt}
\begin{eqnarray}
\lefteqn{
\frac{d^3 N}{dA\,dt\,dE} \Big|_{E=m_{\tilde\tau}}
\Delta A \, \Delta E }
\nonumber\\
&\approx& \frac{10^{-2} \km^{-2}\text{yr}^{-1}}{100\GeV} \:
1\km^2 \:
(2 \cdot 10^{-3} \frac{\GeV}{\cm} \: 1\km)
\nonumber\\
&\approx& 0.02\yr^{-1} \;,
\end{eqnarray}
}%
where the first quantity in the first line is the flux of (single) staus
at the stau rest energy according to Fig.~\ref{fig:SingleSpectra}.
Moreover, $\Delta E$ is the energy loss of
a charged particle in the detector.  Since at small energies it is
dominated by ionisation effects, which are nearly independent of the
particle mass, we can estimate it using
$|dE/dx| \approx 2 \cdot 10^{-3} \GeV\,\text{cm}^2\text{g}^{-1}
 \cdot \rho_\text{ice}$ \cite{Reno:2005si}.
Besides, we have assumed a cubic detector with a volume of $1\km^3$,
which is able to observe the rather low-energy taus from stau decays.
In the case of IceCube, the latter may only be true for the inner core of the
detector \cite{SPIERING}.

\section{CONCLUSIONS}

We have studied the possibility of observing long-lived charged
supersymmetric particles in neutrino telescopes.  If they exist, such
particles can be produced in pairs by cosmic neutrino interactions with
nucleons inside the Earth.  They can then be observed as nearly parallel
tracks of charged particles in the detector, with negligible background
from Standard Model processes.  Unfortunately, the event rates in a
cubic kilometre sized neutrino telescope are rather small if we assume the
Waxman-Bahcall flux for cosmic neutrinos and a SUSY mass spectrum
corresponding to an SPS benchmark point.  However, up to $50$ events per
year are possible, if the neutrino flux is close to its experimental
limit and if the superparticles are comparatively light.

We have also discussed an alternative signal from the decay of
long-lived SUSY
particles that are stopped in the detector.  Even in the most optimistic
case, not more than a few events per decade can be expected in the
detectors currently under construction, so that we have to conclude that
this signal will not be
observable in the near future.

\section*{ACKNOWLEDGEMENTS}
I would like to thank Markus Ahlers and Andreas Ringwald for the
collaboration leading to \cite{Ahlers:2006pf}, on which this talk is
based, Michael Ratz and Christian Spiering for valuable discussions, and
last but not least the organisers of NOW 2006 for the marvellous
workshop.

\providecommand{\bysame}{\leavevmode\hbox to3em{\hrulefill}\thinspace}

\end{document}